\documentclass[%
 reprint,
superscriptaddress,
 amsmath,amssymb,
 aps,
]{revtex4-2}

\usepackage{graphicx}
\usepackage{dcolumn}
\usepackage{bm}
\usepackage[usenames,dvipsnames]{xcolor}
\usepackage{siunitx}
\usepackage[version=4]{mhchem}

\begin{document}

\preprint{APS/123-QED}

\title{Evaporation of Concentrated Polymer Solutions is Insensitive to Relative Humidity}%

\author{Max Huisman}
\affiliation{SUPA and School of Physics and Astronomy, The University of Edinburgh, Peter Guthrie Tait Road, Edinburgh EH9 3FD, United Kingdom}
\author{Paul Digard}
\affiliation{Department of Infection and Immunity, The Roslin Institute, The University of Edinburgh, Easter Bush Campus, Edinburgh EH25 9RG, United Kingdom}
\author{Wilson C. K. Poon}
\affiliation{SUPA and School of Physics and Astronomy, The University of Edinburgh, Peter Guthrie Tait Road, Edinburgh EH9 3FD, United Kingdom}
\author{Simon Titmuss}
\affiliation{SUPA and School of Physics and Astronomy, The University of Edinburgh, Peter Guthrie Tait Road, Edinburgh EH9 3FD, United Kingdom}%

\date{\today}

\begin{abstract}
A recent theory suggests that the evaporation kinetics of macromolecular solutions is insensitive to the ambient relative humidity (RH) due to the formation of a `polarisation layer' of solutes at the air-solution interface. We confirm this insensitivity up to RH~$\approx 80\%$ in the evaporation of polyvinyl alcohol solutions from open-ended capillaries. To explain the observed drop in evaporation rate at higher RH, we need to invoke compressive stresses due to interfacial polymer gelation. Moreover, RH-insensitive evaporation sets in earlier than theory predicts, suggesting a further role for a gelled `skin'. We discuss the relevance of these observations for respiratory virus transmission via aerosols. 
\end{abstract}

\maketitle


`Hindered evaporation' from a water-air interface through some barrier is ubiquitous in applications. For example, as paint or ink dries, a `skin' may form rapidly at the interface~\cite{Tirumkudulu2022}. This affects the `open time' during which a second coat may be applied without imperfections of the first coat showing through~\cite{Holl}. The skin may buckle~\cite{Pauchard2003} or give rise to bubbles~\cite{Arai2012} and diminish coating quality. Such applications have motivated a number of fundamental studies~\cite{Bornside1989,deGennes2002,Doi2006,Arai2012,Raju2022}. 

Hindered evaporation is also important for understanding biological response to environmental relative humidity, RH = $a_e \times 100\%$, where $a_e$ is the water activity in the air.  The evaporation rate of water through the inert wax cuticle of a leaf is proportional to $(1-a_e)$,  but is humidity independent in human skin for RH~ $\lesssim 85\%$~\cite{Roger2016}. It is suggested that the phase behavior of the lipid mixture in the stratum corneum, the outer layer of skin, turns it into an active barrier~\cite{Sparr2001} that responds to changes in $a_e$ to maintain `evaporative homeostasis'. The observation of liquid crystallinity at the water-air interface of lipid solutions displaying RH-independent evaporation rate up to RH~$\approx 80\%$~\cite{Roger2016} supports this picture. 

A recent theory~\cite{Salmon2017} suggests that RH-independent evaporation does not require such `active' response, but occurs whenever a concentrated `polarisation layer' of solutes builds up at the water-air interface due to the advective flux towards the interface driven by evaporation balanced by an opposite diffusive flux. The evaporation rate is primarily controlled by $a(\varphi_i)$, the water activity at the interface with solute volume
fraction $\varphi_i$. If $a(\varphi)$ drops sharply enough at high $\varphi$, the evaporation rate becomes RH insensitive. This effect is enhanced if the mutual diffusivity of the solution $D(\varphi)$ has a similar dependence on $\varphi$. Since such $a(\varphi)$ and $D(\varphi)$ occur widely in macromolecular solutions, RH-insensitive evaporation should be generic even without `active' polarisation layers. Using measured $a(\varphi)$ and $D(\varphi)$, Salmon et al.~predict RH-independent evaporation of polyvinyl alcohol (PVA) solutions up to near saturation (RH $\to 100\%$). 

There are multiple motivations for verifying this theory. Fundamentally, it is important to know whether there is indeed a generic mechanism for RH-insensitive evaporation, and, if so, what active response~\cite{Sparr2001} may add. Applications to coatings will also benefit from a verified predictive theory for the effect of RH. 

More topically, the respiratory droplets that transmit SARS-CoV2 and other pathogens contain a mixture of salt, lipids and glycoproteins (mucins), the latter at high concentrations deep inside the lungs~\cite{Nicas2005}. Their evaporative kinetics controls air-borne transmission~\cite{Wells1934,Xie2007,Netz2020,Netz2021,Dbouk2020}.  Previous empirical studies suggest an intriguing non-monotonic dependence of viral viability on RH \cite{Huynh2022}. A meta-analysis correlates SARS-CoV2 infectivity with RH but provides no clear picture~\cite{Mecenas2020}. As evaporative kinetics probably plays a role in explaining seasonal variability of respiratory virus transmission \cite{ALI2022, Baumgartner2012}, recent studies have considered macromolecular effects on the RH dependence of droplet drying and virus viability~\cite{Poon2020,Merhi22,Pease21}.

\begin{figure*}[t!]
\includegraphics[width=0.82\textwidth]{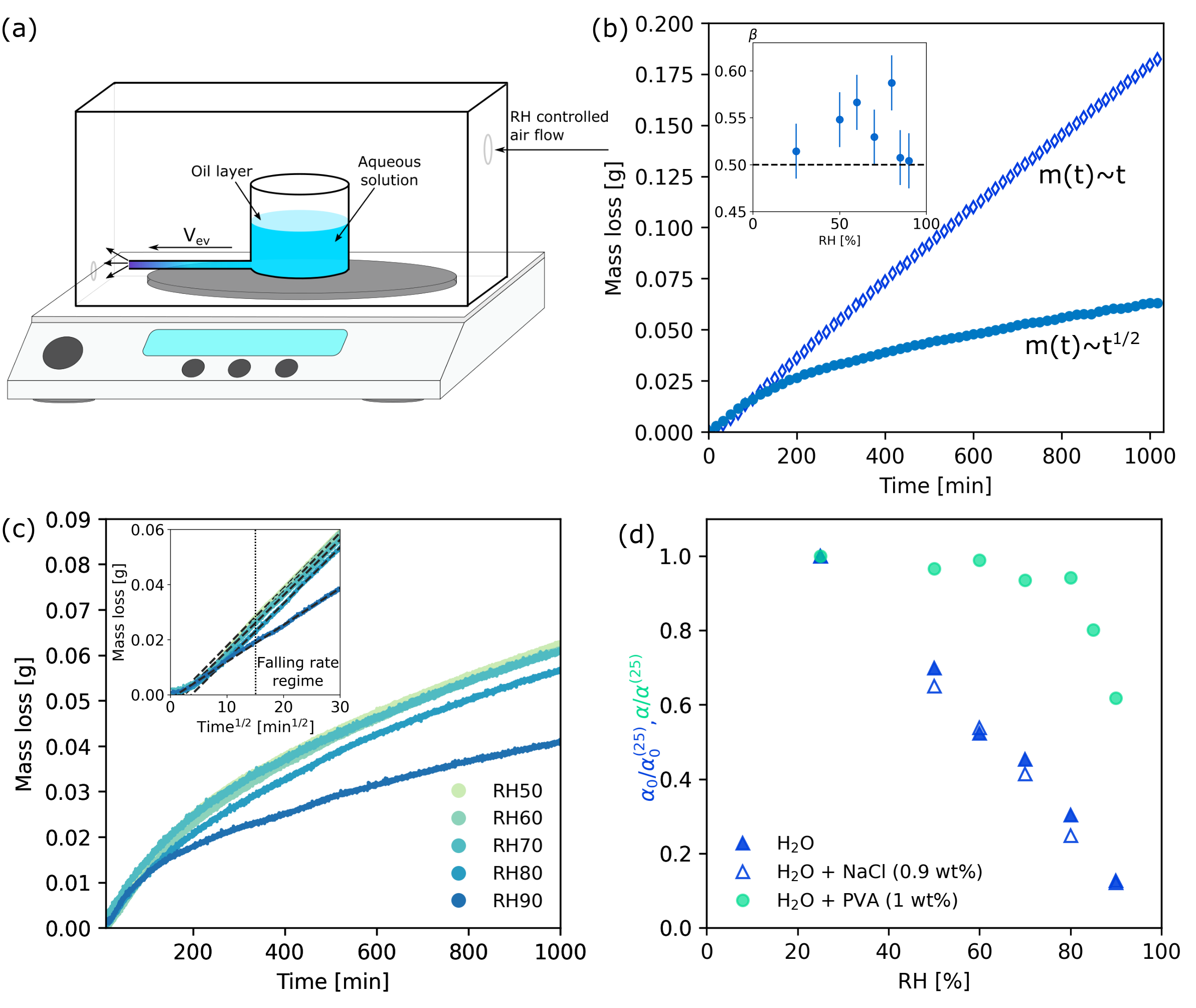}
\caption{\label{fig:mass_loss} (a) Experimental schematic showing one of the five \SI{50}{\milli\meter} capillaries emerging from a \SI{30}{\milli\meter}  section of \SI{20}{\milli\meter} diameter perspex tubing glued to a microscope slide. The 5 capillaries increase evaporation rate, thus allowing faster experiments. (b) Mass loss, $m(t)$, of pure water (\ce{H2O}, open diamonds) and \ce{H2O} + PVA at $\varphi = 0.008$ (filled circles) solutions at RH~$=50$\% and $T=\SI{291}{\kelvin}$. Inset: time exponents $\beta$ of a power-law fit to the late-time $m(t)$ for PVA solutions at different RH. (c) PVA $m(t)$ at different RH plotted against time $t$. Inset PVA $m(t)$ versus $t^{1/2}$. Dashed lines are linear fits at long times, after a transient period indicated by the dotted line where $t>300~$min or $t^{1/2}>15~$min$^{1/2}$. (d) Shaded triangle represents the mass loss rate of pure water, $\alpha_0 = \dot m(t)$, normalised to the value at RH~=~25\%; Open triangle represents the same quantity for 0.9\% (w/w) \ce{NaCl} solutions; Shaded circle represents the slope of the linear portions in the inset of part (c) for $\varphi = 0.008$ PVA solutions, ${\rm d}m(t)/{\rm d}t^{1/2} = (2\sqrt{D_0}\rho A)\alpha$, normalised to the value at RH~=~25\%.}
\end{figure*}

Motivated by these reasons, we have tested experimentally the theory of Salmon et al.~for RH-insensitive hindered evaporation~\cite{Salmon2017}. One of the  geometries they treated,  unidirectional drying from one end of a capillary whose other end is connected to a constant solute concentration reservoir, was previously used to study the drying of lipidic~\cite{Roger2016,Roger2018, Roger2021} and saliva-containing solutions~\cite{Merhi22}. They made testable predictions for PVA solutions up to a single numerical prefactor. 

Based on a previous set-up~\cite{Roger2016,Roger2018,Andersson2018}, Fig.~\ref{fig:mass_loss}(a), we connected 5 rectangular borosilicate capillaries ($\SI{0.20}{\milli\meter} \times \SI{4}{\milli\meter}$, VitroTubes) to a liquid reservoir. To ensure evaporation only at the open end of each capillary, we covered the liquid in the reservoir with a thin layer of 1-Octadecene (Sigma Aldrich). The set-up rested on a Sartorious Secura 224-1S high-precision scale to quantify evaporative mass loss. A sealed enclosure over the sample connected to a Cellkraft P-10A humidifier controlled the RH and temperature $T$, with set-points confirmed using external probes. This obviates the need for blowing constant-RH air at the interface~\cite{Roger2016}, which may perturb drying. The humidifier airflow was tuned to minimize disturbance on mass  measurements. 

A drop shape analyser (Kruss EasyDrop DSA20E) observes the side view at the open end of the capillary. We adjusted the reservoir level to obtain an initial flat water-air interface. The degree of flatness did not visibly change over an experiment, thus minimising curvature effects~\cite{Thomson1872}. The mass loss rate of water should then follow
\begin{equation}
    \dot m(t) = kAc_{\rm{sat}}(1-a_{e}) \stackrel{\rm def}{=} \alpha_0, \label{eq:linear}
\end{equation}
with $A$ the capillary's cross section, and $k$ and $c_{\rm{sat}}$ the mass transfer coefficient and the saturation concentration of water in air~\cite{cussler1997diffusion}. Indeed, $m(t)$ is linear at RH~=~50\%, Fig.~\ref{fig:mass_loss}(b), and  $\alpha_0$ is strictly proportional to $(1-a_{e})$~\cite{SI}.

At early times, the evaporation of a polymer solution with volume fraction $\varphi \ll 1$ should follow Eq.~\ref{eq:linear}. Once a significant polarisation layer forms, the reduced interfacial water activity, $a(\varphi_i)$, controls the evaporation rate, which now reads $\dot m(t) = kAc_{\rm{sat}}(a(\varphi_i)-a_{e})$. As the polarisation layer grows, $\varphi_i$ increases asymptotically towards $\varphi^*$ given by $a(\varphi^*) = a_{e}$, and the evaporation rate becomes sub-linear as $a(\varphi_i) \to a(\varphi^*)$. Solving the case for constant $D(\varphi) = D_0$, the single-coil diffusivity, to highlight the role of water activity, Salmon et al.~predict 
\begin{equation}
    \dot m(t) = f \rho A \sqrt{\frac{\varphi^* D_0}{2\varphi_0 t}} \stackrel{\rm def}{=} \alpha \rho A \sqrt{\frac{D_0}{t}}, \label{eq:squareroot}
\end{equation}
where $f$ is a numerical constant of order unity, $\rho \approx \SI{998}{\kilo\gram\meter^{-3}}$ is the mass density of the water and $\varphi_0$ is the reservoir concentration. At \SI{298}{\kelvin}, 99\% hydrolyzed PVA with molecular weight $M_{\rm{W}}=$~85k-124k has a hydrodynamic radius $R_{\rm{H}}=10.5 \pm \SI{0.5}{\nano\meter}$~\cite{Perfetti2020}. For our PVA with about 50\% higher $M_{\rm w}$, $R_{\rm{H}} \approx \sqrt{1.5} \times 10.5~\rm{nm} \approx 13~\rm{nm}$, from which we estimate $D_0 \approx \SI{16}{\micro\meter^{2}\second^{-1}}$. Equation~\ref{eq:squareroot} also defines a dimensionless $\alpha \propto {\rm d}m/{\rm d}t^{1/2}$ (cf.~Fig.~\ref{fig:mass_loss}(c) inset) to facilitate comparison with data, 
\begin{equation}
    \alpha = (2\rho A \sqrt{D_0})^{-1} \frac{{\rm d}m}{{\rm d}t^{1/2}} = f\sqrt{\frac{\varphi^*}{2\varphi_0}}. \label{eq:prefactor}
\end{equation}

We dissolved PVA ($M_{\rm w}=$~146k-186k, 99\% hydrolized, Sigma Aldrich) in milli-Q water. Figure \ref{fig:mass_loss}(b) shows $m(t)$ for $\varphi = 0.008$  ($= 1\%$ (w/w)\footnote{We use a mass density of PVA, $\rho_{\mathrm{PVA}}=1250\,\mathrm{kgm^{-3}}$ correpsonding to the middle of the 1190-1310\,$\mathrm{kgm^{-3}}$ range given at https://pubchem.ncbi.nlm.nih.gov/compound/11199 }) at $T = \SI{291}{\kelvin}$ and RH~=~50\%. An initially linear regime of constant evaporation rate becomes sub-linear as time progresses. For each RH studied, Fig.~\ref{fig:mass_loss}(c), the sub-linear `falling rate' regime is consistent with a $t^{1/2}$ scaling (inset)~\cite{Salmon2017}. 

\begin{figure}[t]
\includegraphics[width=0.45\textwidth]{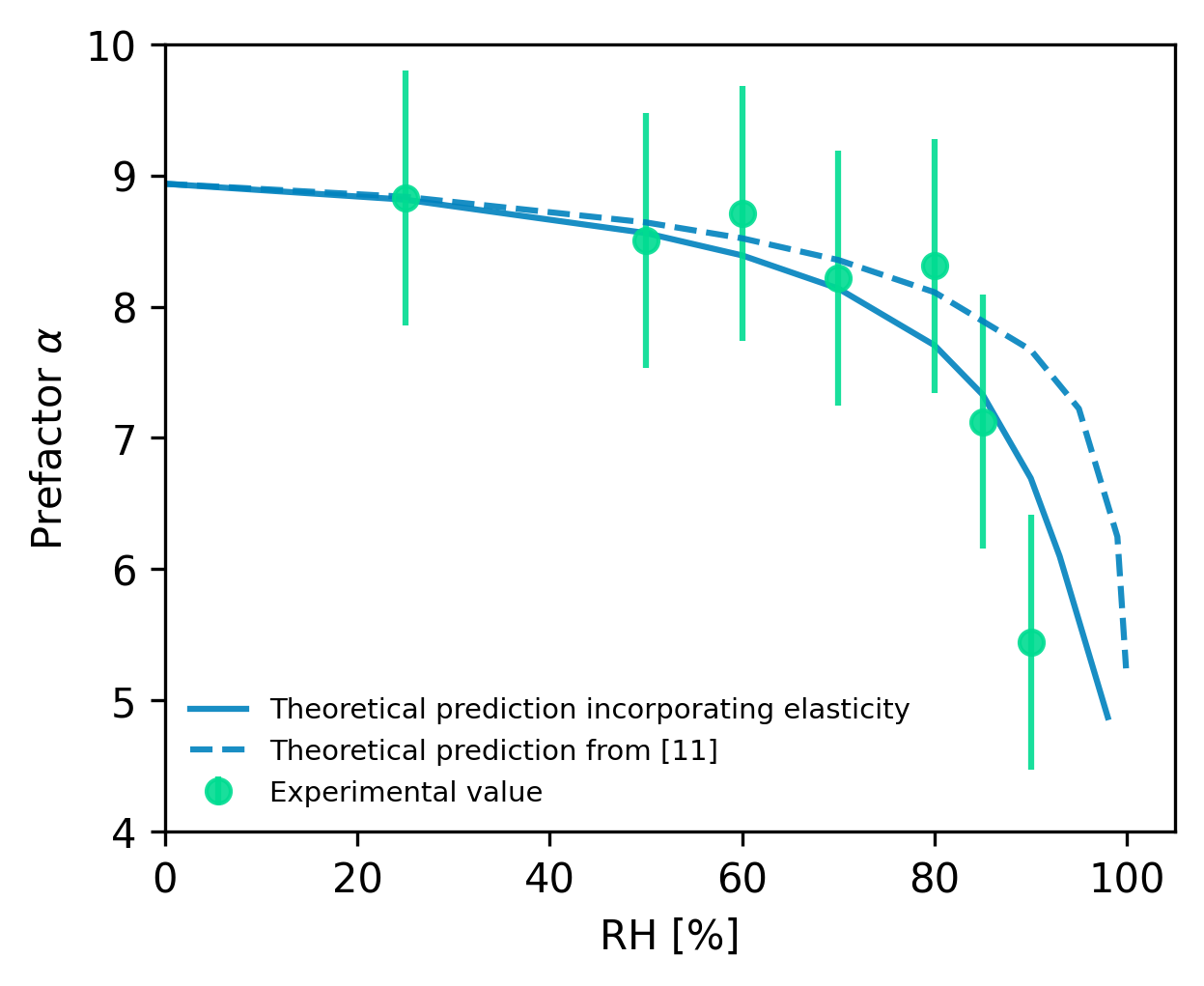}%
\caption{\label{fig:theory} Comparison of the measured evaporation prefactor $\alpha$ as a function of RH (Shaded circle, with error bars giving the standard deviation) with theoretical predictions \cite{Salmon2017} for constant polymer diffusivity and a water activity $a(\varphi)$ without elasticity, Eq.~\ref{eq:a}  (dashed line), and with elasticity, Eq.~\ref{eq:activity} (full line).}
\end{figure}

To highlight the near-RH independence up to 80\% and contrast this with the behavior of pure water, Fig.~\ref{fig:mass_loss}(d) plots $\alpha_0/\alpha_0^{(25)}$ and $\alpha/\alpha^{(25)}$ against RH, where the denominators are the respective values at RH = 25\%~\footnote{While $\alpha^\prime = \dot m(t)$ is the evaporation rate for \ce{H2O}, it is incorrect to call $\alpha \propto {\rm d m}/{\rm d} t^{1/2}$ an `evaporation rate' for  PVA (or lipid \cite{Roger2016}) solutions: see Eq.~\ref{eq:squareroot}.}. This difference is not seen for  0.9\% (w/w) \ce{NaCl} solutions, Fig.~\ref{fig:mass_loss}(d) [$\triangle$]: the macromolecular nature of PVA is key. Moreover, no ordering is needed: crossed-polar observation of the open end of capillaries with evaporating PVA solution showed no liquid crystallinity. 

To test the theory further, consider the absolute value of the evaporative prefactor, $\alpha$, in the falling rate regime. Experimental values calculated using measured ${\rm d} m/{\rm d}t^{1/2}$ and known $(\rho, A, D_0)$ as input, Eq.~\ref{eq:prefactor}, are shown in Fig.~\ref{fig:theory} ({\color{green} $\bullet$}). 
To compare with theory, we solved $a(\varphi^*) = a_e$ to give $\varphi^* = a^{-1}(a_e)$ using a  parameterisation of the experimentally measured $a(\varphi)$~\cite{Jeck2011,Salmon2017}
\begin{equation}
a(\varphi) = (1-\varphi) \exp{[\varphi + \chi \varphi^2]}, \label{eq:a}
\end{equation}
with the Flory-Huggins parameter 
\begin{equation}
       \chi(\varphi) = 3.94 - 3.42(1-\varphi)^{0.09}. \label{eq:chi}
\end{equation}
This `cliff-like' $a(\varphi)$, Eq.~\ref{eq:a}, means $\varphi^* = a^{-1}(a_e)$ varies little for $0 < \rm RH \lesssim 95\%$, engendering even lower RH-dependence for $\alpha \propto \sqrt{\varphi^*}$, Eq.~\ref{eq:prefactor}. Our predicted $\varphi^*$ and a prefactor of $f = 1.12$, Eq.~\ref{eq:prefactor}, account quantitatively for the observed $\alpha( \rm RH)$ up to RH~$\approx 80\%$, Fig.~\ref{fig:theory} (dashed line); discrepancies at higher RH require explanation.

\begin{figure}[t]
\includegraphics[width=0.4\textwidth]{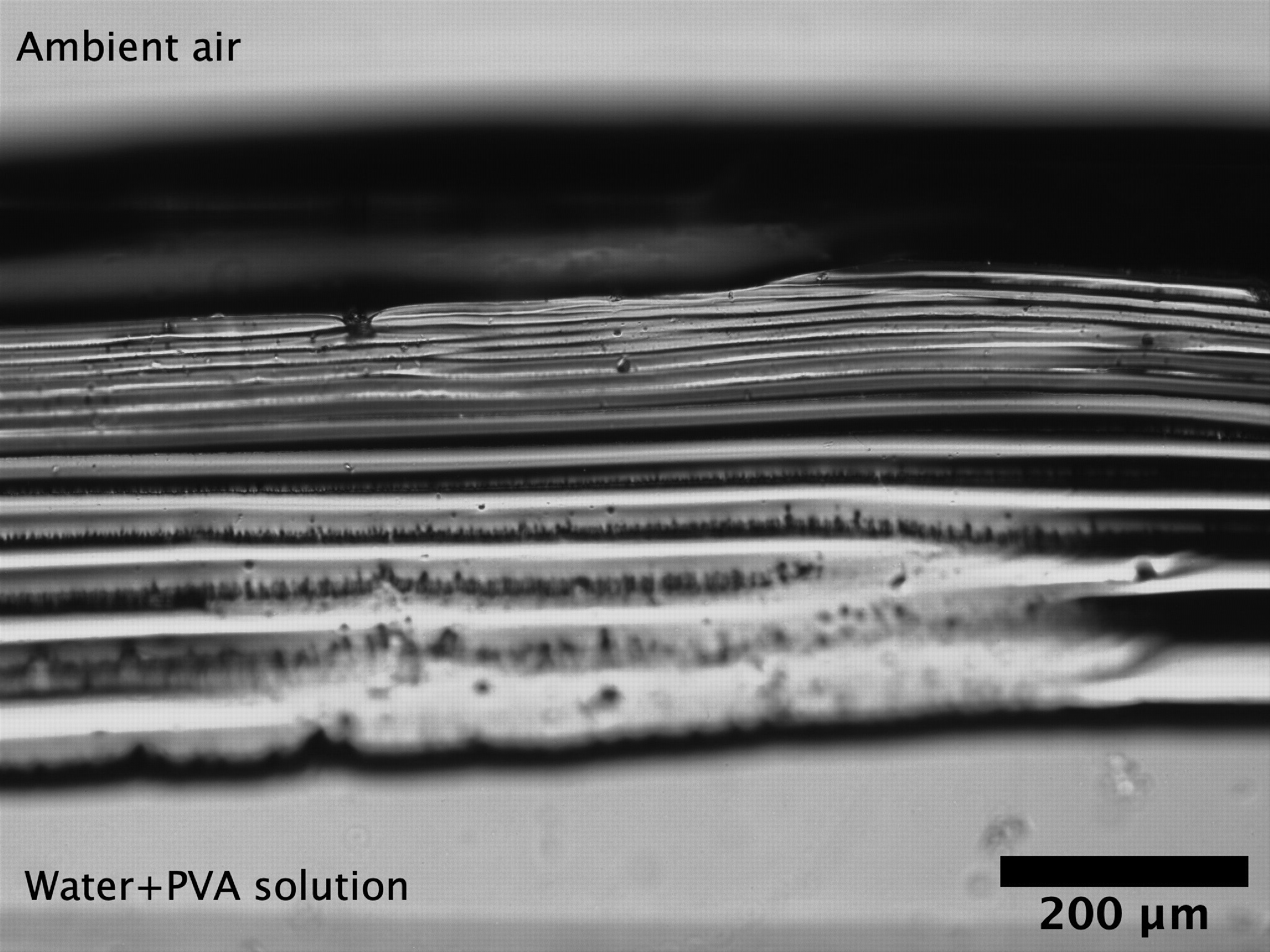}
\caption{\label{fig:exponent} Bright-field image of the polarization layer at the water-air interface at the end ($t\sim10^{3}~$min) of a typical experiment at RH$=50\%$.}
\end{figure}

Bright-field imaging of the polarisation layer, Fig.~\ref{fig:exponent}, shows late-stage delamination consistent with quasi-periodic buckling, allowing air ingress to give bright-field contrast. Such buckling is typical of a stiff film atop a compliant substrate when the film is under considerable compressive stress~\cite{Gao2012}. 
In our case, we have a stiff gel skin~\cite{Okuzono2006, Ozawa_2006,Tirumkudulu2022} covering a more compliant, viscoelastic polarisation layer. The resulting stress contributes to the osmotic pressure~\cite{Leibler1993} and modifies Eq.~\ref{eq:a} to
\begin{equation}
    a(\varphi) = (1-\varphi) \exp{\left[\varphi + \chi \varphi^2 - \frac{K_{\rm{g}}\nu_1}{k_{\rm{B}}T} \ln{\left(\frac{\varphi}{\varphi_g}\right)}\right]},
    \label{eq:activity}
\end{equation}
with $\nu_1$ the molecular volume of water and $K_{\rm g}$ the osmotic modulus of the gelled skin~\cite{SI}. Equation~\ref{eq:activity} well fits our data up to RH $\approx 90\%$, Fig.~\ref{fig:theory} (solid line), with $K_{\rm{g}} = 10\pm \SI{1}{\mega\pascal}$ and $\varphi_g= 0.24\pm 0.02$ as fit parameters.

As already noted, the apparently linear late time ($t \gtrsim \SI{300}{\minute}$) data in Fig.~\ref{fig:mass_loss}(c) [inset] seems consistent with $m(t) \sim t^{0.5}$. Fitting this data to $m(t) \sim t^\beta$ gives $0.5 < \beta < 0.6$, with no systematic dependence on RH, Fig.~\ref{fig:mass_loss}(b) [inset]: evaporation is  always somewhat faster than theory~\cite{Salmon2017} predicts, probably due to the parallel pathways provided by air ingress following the buckling delamination of the polymer skin from the glass wall, giving higher permeability than an intact polarisation layer.

\begin{figure}[t]
\includegraphics[width=0.45\textwidth]{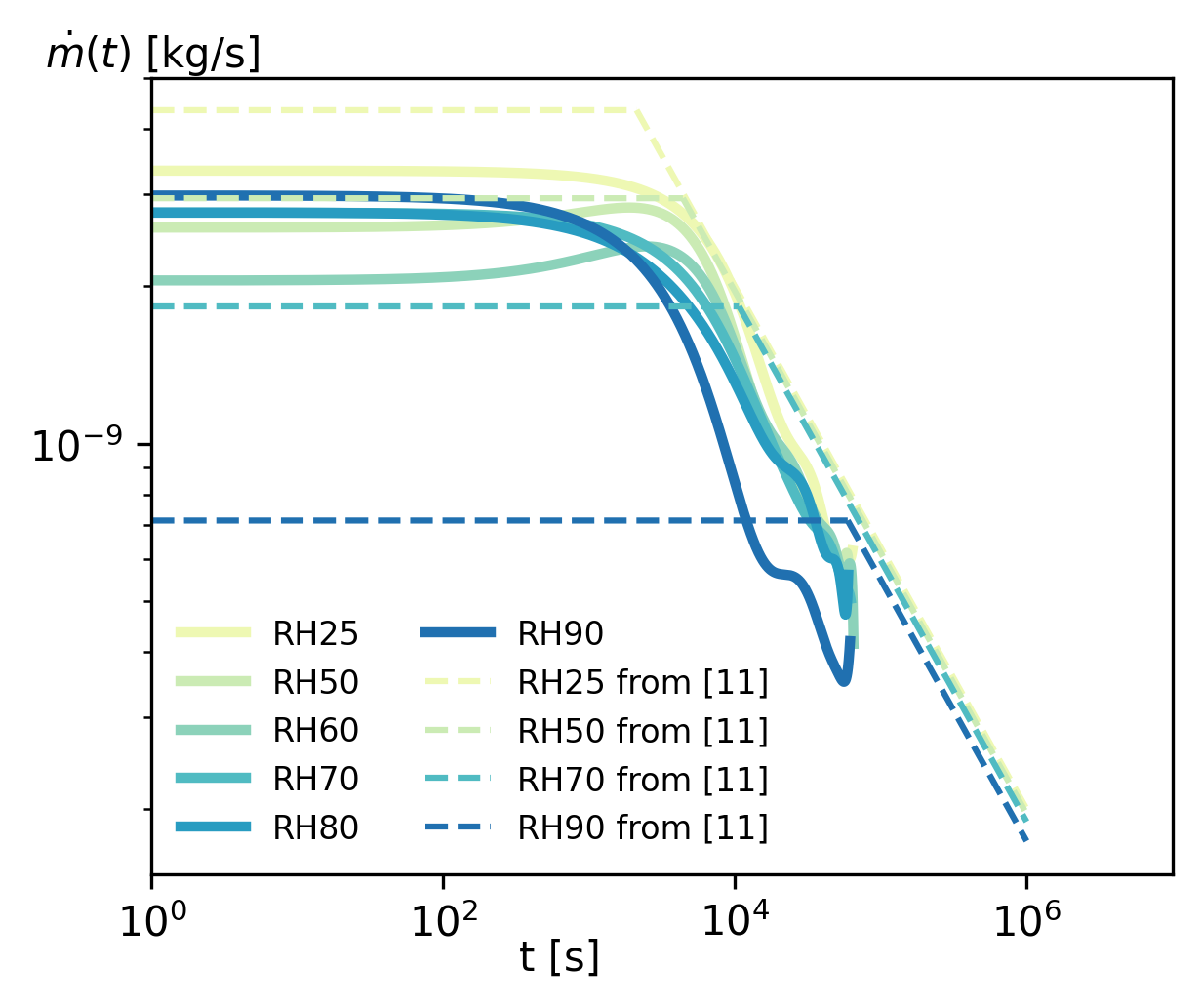}
\caption{\label{fig:loglograte} The dependence of evaporation rate on time plotted on log-log scales. Dotted lines represent for early times the the dilute evaporation regime, Eq. \ref{eq:linear}, and for long times the diffusion limited evaporation regime, Eq. \ref{eq:squareroot} (with $f=1$), the transition point is taken as the intercept.}
\end{figure}

For a final test of theory, we show $\dot m(t)$ in Fig.~\ref{fig:loglograte}. Salmon et al.~\cite{Salmon2017} predict that, before a polarisation layer is established, $\dot m(t)$ should be constant, $\sim t^0$, and essentially that of pure water and decreases linearly with RH, Eq.~\ref{eq:linear}. Including the nearly-universal $t^{-1/2}$ scaling, Eq.~\ref{eq:squareroot} with $f=1$, on the same plot defines a critical time, $t_c$, for the transition between these two regimes. Measured $\dot m(t)$ (full lines) and theory (dashed lines) disagree on two accounts. First, rather than a systematic decrease with RH, the constant $\dot m$ at short times appears to vary randomly with RH. This scatter is comparable to the reproducibility between runs at a single RH, suggesting that $\dot m$ may be RH independent at early times. Secondly, the observed cross-over from $t^0$ to $t^{-1/2}$ scaling occurs at a time that is randomly scattered about $t \lesssim \SI{e4}{\second}$, while the theoretical `knee' systematically shifts to longer times as RH increases. 

These discrepancies motivate the search for extra physics beyond Salmon et al.~\cite{Salmon2017}, which may again relate to a gelled polymer skin at the air-solution interface. Its formation, in $\lesssim \SI{1}{\second}$ in our case~\cite{SI}, gives rise to an interfacial concentration that remains close to $\varphi_g$ for some time~\cite{Ozawa_2006,Okuzono2006,Okuzono2008}. In this `skin-limited' period, the water flux into the skin from the solution is set by the pressure gradient at the skin~\cite{Punati_2022}, and therefore by $\varphi=\varphi_g$, rather than $\varphi^*= a^{-1}(a_e)$~\cite{Salmon2017}. This predicts RH-independent, constant evaporation rate at early times, consistent with the observed random scatter of $\dot m$ values.

The rate at which the interface concentration increases from $\varphi_g$ to $\varphi^*$ is set by an RH-independent skin-limited evaporation rate; moreover, we have seen that $\varphi^*$ is RH insensitive. So, the duration of the skin-limited period, $t_c$, is also insensitive to RH, as observed.

In summary, a PVA-water mixture is found to evaporate almost independently of the ambient RH during early constant evaporation rate and late falling-rate regimes. The latter agrees with a theoretical model~\cite{Salmon2017} predicated on the formation of a polymer polarization layer at the air-water interface driven by advection.  We find evidence that the interfacial gelation of the PVA contributes to the early-stage RH-insensitivity. These findings may prove significant for the evaporation of pathogen-containing respiratory droplets, which include surfactants and gel-forming macromolecules that have been shown to encapsulate dried droplets~\cite{Vejerano2018,Huynh2022}. Our tractable experimental system provides a baseline for investigating the role of these more complex solutes in viral transmission via respiratory droplets~\cite{ZANIN2016159}.

We thank Paul Harris and Andrew Schofield for technical assistance and Jean-Baptiste Salmon, Jay Marsden, Veronica McKinny, Davide Marenduzzo and Patrick Warren for insightful discussions. Aidan Brown lent us a humidifier purchased under EPSRC EP/S001255/1. MH was funded by the University of Edinburgh. 

For the purpose of open access, the author has applied a Creative Commons Attribution (CC BY) licence to any Author Accepted Manuscript version arising from this submission.

\end{document}


\preprint{APS/123-QED}

\title{The evaporation of concentrated polymer solutions is insensitive to relative humidity - Supplementary Information}%

\author{Max Huisman}
\affiliation{SUPA and School of Physics and Astronomy, The University of Edinburgh, Peter Guthrie Tait Road, Edinburgh EH9 3FD, United Kingdom}
\author{Paul Digard}
\affiliation{Department of Infection and Immunity, The Roslin Institute, The University of Edinburgh, Easter Bush Campus, Edinburgh EH25 9RG, United Kingdom}
\author{Wilson C. K. Poon}
\affiliation{SUPA and School of Physics and Astronomy, The University of Edinburgh, Peter Guthrie Tait Road, Edinburgh EH9 3FD, United Kingdom}
\author{Simon Titmuss}
\affiliation{SUPA and School of Physics and Astronomy, The University of Edinburgh, Peter Guthrie Tait Road, Edinburgh EH9 3FD, United Kingdom}%

\date{\today}

\maketitle

\section{Effect of a gelled polymer skin on interfacial activity of water}
The delamination of a solid-like film from the glass-solution interface on the inside of the capillary shown in Figure 3(b) is a signature of the buckling that occurs when the stress in a stiffer thin film, such as the crust that can form at the surface of a drying polymer solution~\cite{Ozawa_2006}, on a more compliant subphase (the underlying polymer solution), exceeds a critical value (see \cite{Gao2012} for a review).

This implies that there will be an interfacial elastic stress contribution to the free energy of the system that is not included in the theoretical approach of Salmon~\cite{Salmon2017}. We suggest that the elastic film corresponds to the solid skin at the surface of evaporating polymer solutions considered by Doi and co-workers~\cite{Okuzono2008,Ozawa_2006}. We follow the approach they use in \cite{Okuzono2008} to calculate the additional elastic  contribution to the osmotic pressure, $\pi_{el}$, of the solution and thus the chemical potential of the water molecules at the air/PVA interface at the open-end of the capillary. This leads to a modification to the functional form of $a(\varphi)$ used to determine $\varphi^{\ast}$ and thus the theoretical prediction of $\alpha$ shown in Figure~2.

Following the approach of Leibler and Sekimoto~\cite{Leibler1993}
\begin{equation}\label{eq:LS}
\pi_{el}(\varphi) = K_{g}\ln{\frac{\varphi}{\varphi_g}}
\end{equation}
where, $K_g$ is the osmotic modulus of the gelled skin layer and $\varphi_g$ is the volume fraction at which the the gelled elastic skin forms. This can be added to the Flory-Huggins mixing contribution,
\begin{equation}\label{eq:FH}
       \pi_{FH}(\varphi) = \frac{k_{\rm{B}}T}{\nu_1}[\frac{\varphi}{N}-\ln{(1-\varphi)} + \varphi + \chi \varphi^2]
\end{equation}
to give,

\begin{equation}
       \pi(\varphi) = -\frac{k_{\rm{B}}T}{\nu_1}[\ln{(1-\varphi)} + \varphi + \chi \varphi^2] + K_{\rm{g}} \ln{\frac{\varphi}{\varphi_g}},
\end{equation}
where, $\nu_1$ is the molecular volume of water and the degree of polymerization $N$ is large, as is the case here, allowing the leading term in equation~\ref{eq:FH} to be neglected.

Using $\mu_1 (\varphi) = -\nu_1 \pi(\varphi)$  and $a(\varphi)= \exp{\left[\mu_1(\varphi)/k_{\rm{B}}T \right]}$ the interfacial water activity is given by,

\begin{equation}
    a(\varphi) = (1-\varphi) \exp{\left[\varphi + \chi \varphi^2 - \frac{K_{\rm{g}}\nu_1}{k_{\rm{B}}T} \ln{\frac{\varphi}{\varphi_g}}\right]}.
    \label{eq:activity}
\end{equation}
We note that although Eq.~\ref{eq:LS} was originally used in the treatment of vapour sorption by glassy polymers~\cite{Leibler1993}, we have used it in the same way as Okuzono and Doi~\cite{Okuzono2008} to provide the simplest functional description for an elastic contribution to the free energy of the gelled `skin' as a whole (and neglecting the likely-different behaviour of the top few molecular layers).
In common with Doi and co-workers~\cite{Doi2006}, we do not attempt to make a  distinction between whether the skin that is responsible for the elastic contribution to the free energy is technically a gel or technically glassy.

\section{Rapid formation of a gelled layer at the interface of an evaporating polymer solution}
\begin{figure}[h]
\includegraphics[width=0.45\textwidth]{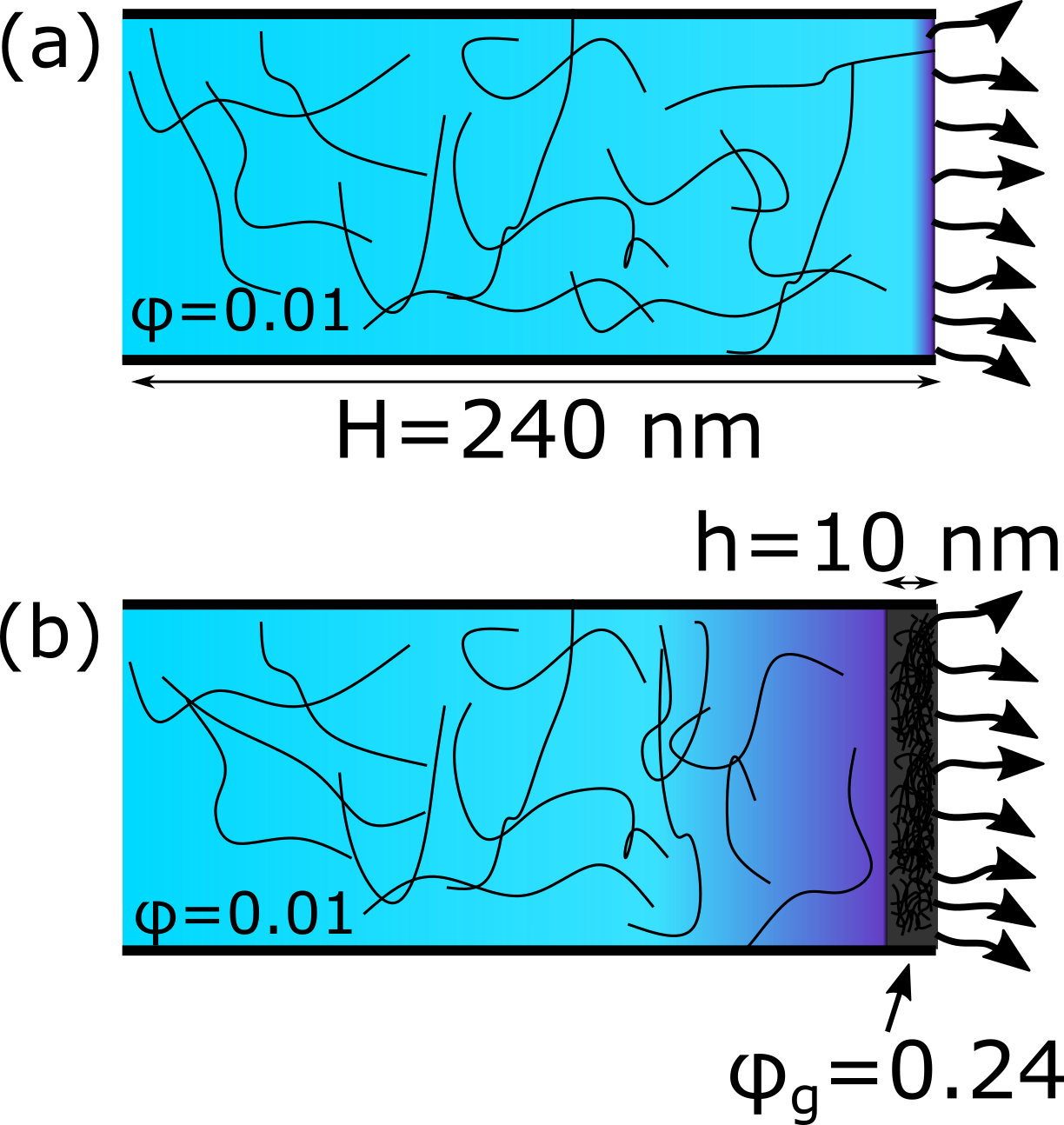}

\caption{\label{fig:gel_skin_rapid} Rapid formation of a gelled layer of thickness 10~nm at $\varphi=\varphi_g=0.24$ from a solution at $\varphi_V=0.01$}
\end{figure}
The formation of an interfacial layer of thickness, $h=10$~nm at $\varphi=\varphi_g=0.24$ from a bulk solution of $\varphi=0.01$ requires the removal of the water by evaporation from a slice of thickness $H=240$~nm. For a capillary of cross-section 0.2~mm$\times$ 4~mm, this corresponds to a mass loss of 0.2~$\mu$g. From Fig. 1(c), the initial mass loss rate from a single capillary containing 1\% PVA solution is 0.6~$\mu$g/s, meaning that a 10~nm thick gelled layer can form within 1~s.